\documentstyle[epsf,12pt]{article}

\newcommand{\be}{\begin{equation}}
\newcommand{\ee}{\end{equation}}

\newcommand{\ind}[1]{{\mbox{\scriptsize #1}}}

\begin{document}
\setlength{\unitlength}{1mm}

\title{Gravitational Capture of Cosmic Strings by a Black Hole
\footnote{Preprint Alberta Thy 14-97}}
\author{\\
Jean-Pierre De Villiers\footnote{e-mail:
jpd@phys.ualberta.ca}  ${}^{1}$, and
Valeri Frolov\footnote{e-mail: frolov@phys.ualberta.ca} ${}^{1,2,3}$
}
\maketitle
\noindent
$^{1}${ \em
Theoretical Physics Institute, Department of Physics, \ University of
Alberta, \\ Edmonton, Canada T6G 2J1}
\\ $^{2}${\em CIAR Cosmology Program}
\\ $^{3}${\em P.N.Lebedev Physics Institute, Leninskii Prospect 53,
Moscow
117924, Russia}

\bigskip

\begin{abstract}
The gravitational interaction of an infinitely long cosmic string with
a Schwarzschild black hole is studied. We consider a straight string
that is initially at a great distance and moving at some initial
velocity $v$  ($0 < v < c$) towards the black hole. The equations of
motion of the string  are solved numerically to obtain the dependence
of the capture impact parameter on the initial velocity. 
\end{abstract}

\bigskip

{\it PACS number(s): 04.60.+n, 12.25.+e, 97.60.Lf, 07.05.Tp}

\newpage
\baselineskip=.8cm

\section{Introduction}

\noindent  
Cosmic strings are topologically stable one-dimensional objects  which are
predicted by grand unified theories. Cosmic strings (as well as other
topological defects) may appear during a phase  transition in the early
Universe. (A detailed discussion of cosmic  strings and other topological
defects can be found in Shellard and Vilenkin \cite{ShVi:94}.) The
characteristic thickness of a cosmic string is $\rho_0\sim l_{\ind{Pl}}
(m_{\ind{GUT}}/m_{\ind{Pl}})$ while its mass per unit length  is 
$\mu\sim(m_{\ind{Pl}}/l_{\ind{Pl}})(m_{\ind{GUT}}/m_{\ind{Pl}})^2$.  Here
$m_{\ind{GUT}}$ is the characteristic mass defining the symmetry breaking in
the grand unified theory which is responsible for the cosmic string 
formation,
and $m_{\ind{Pl}}$ and $l_{\ind{Pl}}$ are the Planck mass and length,
respectively. 

For cosmic strings of astrophysical interest the parameter $\rho_0$ is much
less than any other parameters that enter the problem, such as the length of
the string or the size of inhomogeneity of the  gravitational field in 
which it
is moving, so that one can idealize the motion of the string as a
two-dimensional worldsheet  $x^{\mu}=x^{\mu}(\tau,\sigma)$. It can be 
shown  
that the equations  of motion of a cosmic string in this approximation follow
from the Nambu-Goto action (see e.g., Shellard and Vilenkin \cite{ShVi:94}),
\begin{equation}\label{1}
I=-\mu\int
d^2\xi \sqrt{-\mbox{det} (G) }\, , \hspace{.5cm}
G_{A B}=g_{\mu\nu} {\partial {x^{\mu}}\over\partial 
\xi^A} {\partial {x^{\nu}}\over \partial \xi^B}\, ,
\end{equation}
where $g_{\mu\nu}$ is the spacetime  metric, ${\xi}^{A}$ denote the 
worldsheet
coordinates ($A,B=0,1$; $\xi^0=\tau$, $\xi^1=\sigma$), and  $G_{A B}$ is the
induced metric on the worldsheet of the string. A two-dimensional  worldsheet
which provides an extremum to the Nambu-Goto action  is a minimal surface.

We consider the interaction of a cosmic string with a black hole of
mass $M$. For this problem the gravitational field created by the
string can be neglected because the dimensionless parameter which
characterizes the strength of the field $\mu^*=G\mu/c^2$ is negligibly
small (for GUT one has $\mu^*\sim 10^{-6}$). It means that we can
consider the string as a test object moving in the background of a
black hole metric. We also assume that the length of the string $L$ is
much greater than the gravitational radius, $L\gg r_{\ind{g}}=2GM/c^2$,
and use the infinite-string approximation.

There are two possible outcomes for a cosmic string moving from
infinity  towards a black hole: it may pass near the black hole and
escape to the distant region (gravitational scattering), or it may be
captured by the black hole.  The scattering of a string is an inelastic
process since the string absorbs some energy (i.e. the string's
internal energy, connected with the excitations of its degrees of
freedom, changes during the interaction). The capture of a cosmic
string occurs when the string passes sufficiently close to the
gravitational radius. After capture, the string remains attached to the
black hole. For $\mu L\ll M$ the total mass of the string is much
smaller than the mass of the black hole, and the attached string will
move around the black hole while the black hole remains practically at
rest. In the opposite case, $\mu L\gg M$, the black hole will be
accelerated by the string. The characteristic time of this process is
$T\sim vM/(\mu c^2)\sim (v/c)(\mu^* r_{\ind{g}}/c)$. 

Our aim is to study the gravitational interaction of a cosmic string
with a Schwarzschild black hole and to determine the conditions under which
the string is captured. For simplicity, we consider an infinitely long
string which is initially straight (far from the black hole), and
denote its initial velocity by $v$ and its impact parameter by $b$ (see
Figure~1). For a given velocity $v$, capture occurs if $b$ is less than
the so-called capture impact parameter $b_{\ind{capture}}(v)$. By
solving the equations of motion of the string numerically we  obtain
the dependence of the capture impact parameter on the initial velocity
$v$.

\section{Equations of Motion}

In order to use numerical methods it is convenient to deal with a string of
finite size. For this purpose we modify the problem by taking a finite string
and attaching heavy particles  ("monopoles") at its ends (see Figure~1). In
this case, it is easier to use a quadratic form of the string action,
equivalent to the Nambu-Goto action, namely the Polyakov action
\cite{Polyakov81}. The  corresponding action takes the form,
\begin{eqnarray}\label{2}
S\left[{x}^{\mu},{h}_{AB},{\sigma}_{i}\right]& = 
& -\mu \int_{{\tau}_{1}}^{{\tau}_{2}}
{d\tau \int_{{\sigma}_{1}(\tau)}^{{\sigma}_{2}(\tau)}
{d\sigma\,\sqrt{-h} \,h^{AB}\,G_{AB}}}\\
\nonumber & - &
\sum_{i=1}^{2}{{{m}_{i} \over 2}\int_{{\tau}_{1}}^{{\tau}_{2}}
{d\tau\,\,{g}_{\mu \nu}\,
{d X_i^{\mu}(\tau) \over d\tau}\, 
{d X_i^{\nu}(\tau) \over d\tau}}}
\end{eqnarray}
where $h_{AB}$ is the internal metric with determinant $h$, and
$X_i^{\mu}(\tau)=x^{\mu}(\tau, \sigma_i(\tau))$ is a world-line of the
$i$-particle, and $m_{i}$ is its mass. The end points are taken to
represent the mass of a longer (potentially infinite) string lying
outside the region of interest; the motion of these points represents a
boundary condition for the section of the string worldsheet under
study.  This approach allows us to approximate the motion of an
infinite string by considering the case of extremely heavy particles, when
$m_{i}\rightarrow \infty$. Under this assumption, the variation of the action
of the massive end points with respect to $X^{\mu}_{i}(\tau)$
leads to the equations of motion,
\begin{equation}\label{3a}
{d^2 X_i^{\mu}(\tau) \over d{\tau}^2}+\Gamma^{\mu}_{\alpha\beta}{d
X_i^{\alpha}(\tau) \over d\tau}{d X_i^{\beta}(\tau) \over d\tau}=0\, .
\end{equation}

The equations of motion of a string are obtained by varying the action
(\ref{2}) with respect to $x^{\mu}(\tau, \sigma)$.  We fix the freedom
in the choice of internal coordinates $(\tau,\sigma)$  on the
worldsheet of the string by using a conformal gauge so that $h_{AB}$ 
is of the form ${h}_{A B} = h\, \mbox{diag}(-1,1)$. The remaining freedom is
used to put ${\sigma}_{1}=-\pi/2$ and ${\sigma}_{2}=\pi/2$ for the end points. 
In these coordinates
the equations of motion of the string have the form
\begin{eqnarray}\label{3}
{{\partial}^{2}\,{x}^{\mu} \over \partial\,{\tau}^{2}}
-{{\partial}^{2}\,{x}^{\mu} \over \partial\,{\sigma}^{2}}
+{\Gamma}^{\mu}_{\rho \eta}\,
{\partial\,{x}^{\rho} \over \partial\,\tau}\,
{\partial\,{x}^{\eta} \over \partial\,\tau} 
-{\Gamma}^{\mu}_{\rho \eta}\,
{\partial\,{x}^{\rho} \over \partial\,\sigma}\,
{\partial\,{x}^{\eta} \over \partial\,\sigma}
& = & 0\, (-{\pi \over 2} < \sigma < {\pi \over 2})\\
g_{\mu \nu}\,\left({\partial {x^{\mu}}\over \partial {\tau}}\,
{\partial {x^{\nu}}\over \partial {\tau}} + 
{\partial {x^{\mu}}\over \partial {\sigma}}\,
{\partial {x^{\nu}}\over \partial {\sigma}}\right) & =& 0\\
\nonumber
g_{\mu \nu}\,{\partial {x^{\mu}}\over \partial {\tau}}\,
{\partial {x^{\nu}}\over \partial {\sigma}} & =& 0 
\end{eqnarray}
The last two equations are  constraints.
We  use them as checks on the quality of the numerical solutions to
(\ref{3}).

The equations of motion of the string and the end points need to be solved
simultaneously. In order to do this, it is necessary to relate the proper 
time of the string ($\tau$) and the parameter ($\tau$) appearing in
(\ref{3a}). This is done in the process of performing
variation of the action for the finite length of string, which yields
conditions on the boundaries of the string worldsheet. The details of these
calculations are not important, but it follows from matching the
behaviour of the boundaries of the string worldsheet and the massive
particles that these two time-like parameters must be the same, namely
the proper time. So, together with the equations of motion for the string 
(\ref{3}), the condition that the end points of the string move along
geodesics, 
\begin{equation}\label{3c}
x^{\mu}(\tau,-{\pi\over 2})=X^{\mu}_1(\tau)\, ,\quad
x^{\mu}(\tau,{\pi\over 2})=X^{\mu}_2(\tau)\, ,
\end{equation}
gives the boundary conditions for the string equations.

We assume that the ends of the string are moving on initially parallel
geodesics with velocity $v$. Numerical results demonstrate that as soon as the
length $L$ of the string becomes greater than some quantity $L_m(v)$ the
capture parameter $b_{\ind{capture}}(v)$ does not depend on $L$. In other
words, in this regime the massive end points do not alter the dynamics of the
string itself. Their role becomes one of simulating the behaviour of an
infinite string. More will be said on this in the following sections.

\section{Numerical Method}

The equations of motion of the cosmic string have the form of a system
of non-linear wave equations. After some trials with more traditional
approaches, a method based on Von Neumann's discretization for the
linear wave equation was developed. This approach deals with the wave
equation directly as a second-order equation, with a standard finite
differencing for the time derivative and an averaging method for the
spatial derivative that gives rise to an implicit scheme. Von Neumann's
discretization yields a tridiagonal system that is solved
algebraically. The extended version of the method deals with a
block-tridiagonal system of linearized equations that is solved
iteratively. Details on the numerical method can be found in 
\cite{DeVilliers97b}.

In order to have a well-posed problem, we must specify the shape of the
string at some initial time ${\tau}_{0}$, along with normal derivatives
(${\partial}_{\tau}{x}^{\mu}$) everywhere on the string, including the
massive ends, at this initial time. Taken together, these completely
specify the initial/boundary value problem (IBVP).  Since the Von
Neumann method requires two time steps in order to compute the next
(unknown) time step, a completely equivalent way of specifying the IBVP
is to use two initial time steps. Using the property of asymptotic
flatness for the Schwarzschild spacetime, and the fact that analytic
solutions to the equations of motion in Minkowski spacetime are readily
obtained, the IBVP can be specified using analytic expressions.
Although this approach requires that the solution be started well away
from the black hole, it greatly simplifies the process of starting the
Von Neumann solver. Although a cosmic string cannot be straight when
exposed to an inverse-square force, the choice of the initial position
of the string is made so as to minimize the discrepancy between the
analytic (straight) solution and "true" solution in the black hole
spacetime; the constraint equations are a useful guide in this choice.
The initial position must also be sufficiently distant so that the
capture impact parameter is not affected (starting a straight string
solution too close to the black hole could have a marked effect on the
capture impact parameter).

\section{The Capture of Cosmic Strings}

The portion of the cosmic string that passes closest to the black hole will
experience a larger gravitational force than more distant parts of the string.
Because of this, the string will become deformed as it moves through the
spacetime of the black hole. For more distant parts of the string, where the 
gravitational field is weak, it is possible to describe this deformation by
solving the linearized Nambu-Goto equations. In this approximation the
deformation can be treated as a perturbation to a straight string solution and
decomposed into two components: (1) a component that is directed along the
original direction of motion, and (2) a component that is orthogonal to (1) and
to the straight string itself. This analysis shows that, due to the long-range
nature of the gravitational field, the deformation in the direction of motion
(1) can be large. What is important is that the deformation along (2) remains
finite; this deformation is related to the process of string capture. The
deformation of the central part of the sting determines whether or not capture
will occur, and a numerical solution to the equations of motion is required to
make this determination. 

Because the string has tension, which acts to straighten its bent parts, it is
reasonable to expect that it would be more difficult to capture the string than
a test particle moving with the same velocity and impact parameter. In other
words, the capture impact parameter for a string approaching a black hole
should be less than the capture impact parameter for a test particle. 

The results of the numerical calculations are presented in Figure~2.\footnote{
The gravitational capture of strings by Schwarzschild black holes has already
been discussed by Moss and Lonsdale \cite{LoMo:88}. The results shown above do
not agree with these earlier results. The smallest capture impact parameter for
a string quoted by these authors occurs at the ultra-relativistic limit, and
their capture graph does not reproduce features shown in Figure~2.} It shows
log--log graphs of the capture impact parameter (${b}(v)/{r}_{g}$) as a
function of initial velocity. Dashed, dotted, and dashed-dotted lines represent
capture impact parameters for strings of lengths $L= 100r_{g}$, $1000r_g$, and 
$2000r_g$, respectively. The capture curves for the cosmic strings are shown
with error bars. The lower bound  indicates largest impact parameter resulting
in capture, while the upper bound indicates the smallest impact parameter for
which the string escaped. The data for these curves was obtained using
Eddington-Finkelstein ingoing coordinates since it is easy to determine when
the string becomes trapped (capture occurs when the radial coordinate in the
equatorial plane is less than the gravitational radius). 

The graphs show that for a string of finite size the capture impact
parameter depends on both the length and velocity of the string, $b(v,L)$.
The capture curve for the $L = 2000\,{r}_{g}$ string is definitive for
velocities  $v \ge 0.2 c$, meaning that the capture curves for longer
strings are indistinguishable in this velocity range. Probing the
low-velocity behaviour of very long strings is numerically intensive,
so the extension of the capture curves below $v \approx 0.05 c$ has yet
to be carried out.

The solid line represents the capture impact parameter for a
test particle. The gravitational capture of test particles is well
understood: a particle moving at non-relativistic velocities has a
capture impact parameter ${b}(v)/{r}_{g} = 2/v$, while in the 
ultra-relativistic regime ($v \rightarrow c$), the capture impact
parameter is ${b}(v)/{r}_{g} = 3 \sqrt{3}/2$.  The capture impact
parameter for the ultra-relativistic particle is also the smallest
capture impact parameter. 

The capture curve of a cosmic string is different in many ways from
that of a test particle. Unlike a test particle, the smallest capture
impact parameter for a string occurs at some intermediate velocity (e.g
$v \approx 0.1 c$ for the $L = 2000 \,{r}_{g}$ string). The capture
impact parameter increases as $v \rightarrow c$ and $v \rightarrow 0$.
Also, the smallest capture impact parameter decreases with increasing string
length. At a sufficiently low velocity, the capture curve of a (finite)
string intersects the curve for the test particle. The capture impact
parameter for strings is independent of string length for velocities
approaching $c$, even for very short strings. 

The increase in the capture impact parameter for velocities approaching
the speed of light can be understood in terms of the propagation time
of influences along the string. As $v \rightarrow c$, the interaction
time where the string experiences the gravitational pull of the black
hole becomes progressively shorter, and the influence of the black hole
is felt by a shorter and shorter segment of string. It is reasonable to
expect that in the ultimate limit, the string will behave like a
particle and hence have the same capture impact parameter. 

This result illustrates how the internal tension in the string helps
determine its response to the gravitational field of the black hole. In
the extreme relativistic limit, tension plays no role since signals
cannot propagate sufficiently far down the string to help influence the
capture process. At lower velocities, tension plays an increasingly
dominant role and helps the string avoid capture. The smallest capture
impact parameter for the string is far less than that for a test
particle, and occurs away from the ultra-relativistic limit.

The capture curve for $L=100 r_g$ intersects the capture curve for the
test particle. This reflects the fact that for small velocities the
finiteness of the string plays an important role. The
trajectories of the massive particles terminating the strings are
focused by the black hole. When the strings are sufficiently long, this
focusing is negligible. However, the boundary conditions used in the
numerical work force these massive particles to move on geodesics, so
they are subject to the same capture characteristics as free test
particles. This means that the end particles will be trapped by the
black hole when the initial velocity becomes sufficiently small, no
matter how far they are initially from the black hole. Since there is a
string connecting the two particles, its behaviour will be greatly
influenced by a strong focusing of the end points, and hence easily
captured. For this reason, the intersection of particle and string
curves is to be expected for any finite length of string. However, the
initial velocity for which intersection occurs decreases with
increasing string length. In the ultimate limit, where string length
becomes infinite, the two curves should intersect only when $v = 0$.

When $v \rightarrow 0$, the string is expected to exhibit the
characteristics of a stationary string, provided that the characteristic
time for propagation of signals along the string, ${t}_{prop} \sim
L/c$, is much less than the characteristic time for passage by the
black hole ${t}_{pass} \sim {r}_{g}/v$, or by rearranging, the string
is effectively static provided that $v/c \ll {r}_{g}/L$. Stationary
string solutions have been discussed by Frolov {\it et al}
\cite{Frolov89}. Using their results, it is easy to show that the
minimum  radial coordinate of a stationary string of length L whose end
points are located at $({X}_{0},{Y}_{0},\pm L/2)$ is given by,
\begin{equation}\label{4}
{r}_{min} \approx {r}_{g}+
 \left({{L}^{2} + 4\,{Y}_{0}^{2} \over 8\,{r}_{g}} \right)\,
 \left[{\pi \over 2} - \arctan{{L \over 2 {Y}_{0}}} \right]
\end{equation}
Provided ${r}_{min} > {r}_{g}$ the string avoids capture. However, it is
easily seen that ${r}_{min} \rightarrow {r}_{g}$ as $L \rightarrow \infty$,
so that stationary strings of infinite length are always captured. This
confirms the claim made above.

\section{Conclusion}

In this numerical study of the gravitational capture of cosmic strings,
we modelled strings of infinite length using finite strings with
special (geodesic) boundary conditions. These boundary conditions
proved problematic at low velocities if the string length is made too
short, since the trajectory of the massive end points can be greatly
influenced by the black hole under these conditions (the focusing
problem). However, the capture curve is independent of string length
for velocities close to that of light.  Further, a string of infinite
length is always harder to capture than a test particle {\it except} at
the two extremes, $v \rightarrow 0$ and $v \rightarrow c$, where the
two capture curves converge.

Work is currently underway to obtain the capture curves for strings and
Kerr (rotating) black holes.

\vspace{12pt} 
{\bf Acknowledgements}:\ \ This work was partly supported by the
Natural Sciences and Engineering Research Council of Canada.  One of
the authors (V.F.) is grateful to the Killam Trust for its  financial
support. The authors  wish to thank Dr. J.C. Samson, Chair, Dept. of
Physics, University of Alberta,  for  granting access to a Silicon
Graphics Power Challenge parallel computer on which all numerical work
was carried out. 

\begin{figure}\label{f1}
\let\picnaturalsize=N
\def\picsize{10cm}
\def\picfilename{string_cap_jpd_fig1.eps}
\ifx\nopictures Y\else{\ifx\epsfloaded Y\else\input epsf \fi
\let\epsfloaded=Y
\centerline{\ifx\picnaturalsize N\epsfxsize \picsize\fi
\epsfbox{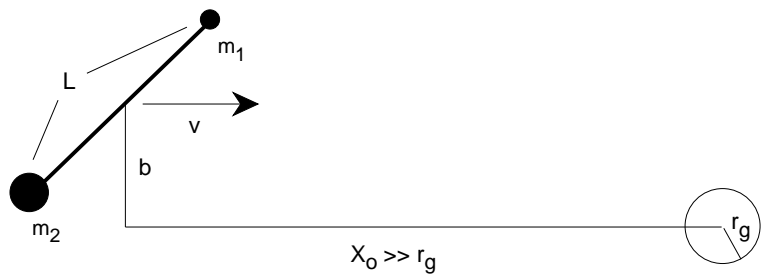}}}\fi
\bigskip
\caption[fig1]{
A straight cosmic string moving in the gravitational field
of a Schwarzschild black hole. Illustration of the scattering problem.
}
\end{figure}

\begin{figure}
\label{f2}
\let\picnaturalsize=Y
\def\picsize{10cm}
\def\picfilename{string_cap_jpd_fig2.eps}
\ifx\nopictures Y\else{\ifx\epsfloaded Y\else\input epsf \fi
\let\epsfloaded=Y
\centerline{\ifx\picnaturalsize N\epsfxsize \picsize\fi
\epsfbox{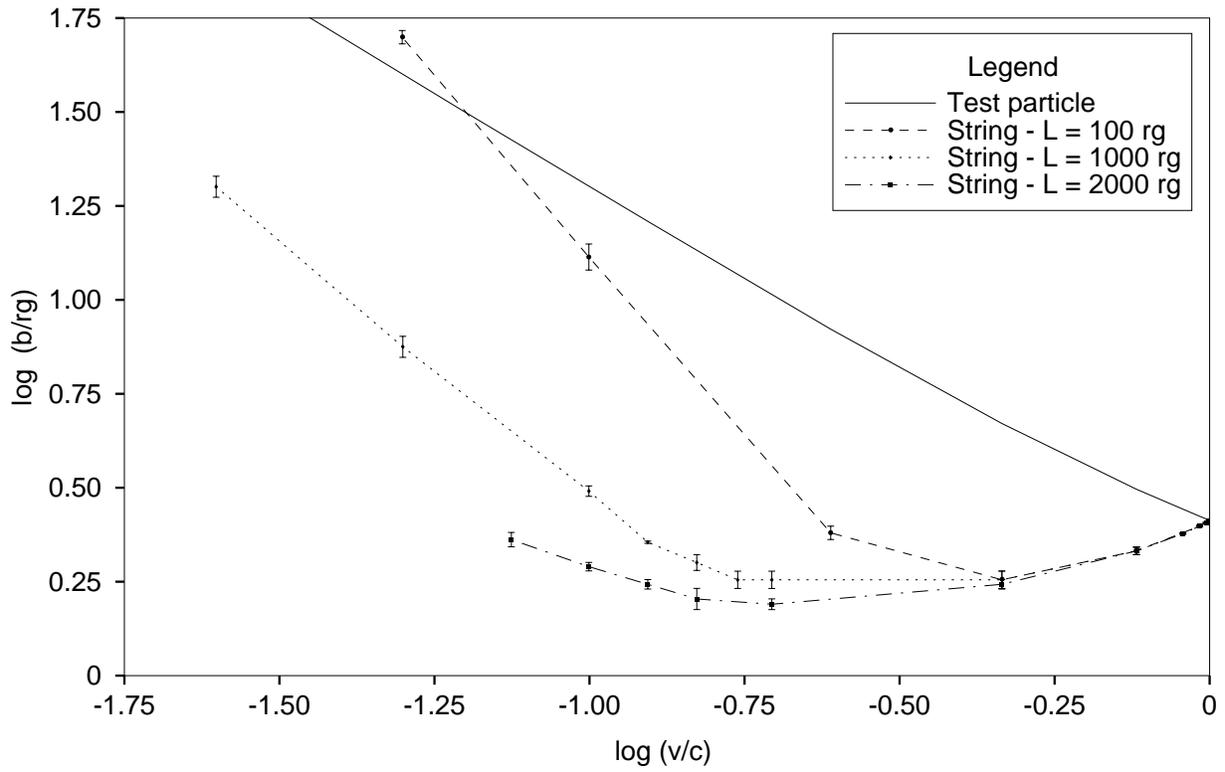}}}\fi
\bigskip
\caption[f2]{The capture impact parameter $b$ as the function of the
initial velocity $v$ of the cosmic string.
}
\end{figure}

\end{document}